\documentclass[pra,onecolumn,floatfix,a4paper,superscriptaddress]{revtex4}
\usepackage{bm,color,graphicx,amsmath}

\usepackage[colorlinks, citecolor=blue,linkcolor=blue]{hyperref}
\usepackage{braket}
\usepackage{amssymb}

\usepackage{float}

\newcommand{\dd}{{\rm d}}
\newcommand{\Tr}{{\rm Tr}}

\newcommand{\D}{\mathcal D}
\newcommand{\I}{\mathrm i}

\begin{document}

\title{Parametrically Enhanced Ergotropy and Charging Power in a Three-Mode Bosonic Quantum Battery}

\author{Mustapha Tarboun}
\affiliation{LPTHE-Department of Physics, Faculty of sciences, Ibnou Zohr University, Agadir, Morocco}	
\author{Mohamed Amazioug} 
\thanks{m.amazioug@uiz.ac.ma}
\affiliation{LPTHE-Department of Physics, Faculty of sciences, Ibnou Zohr University, Agadir, Morocco}	

\date{\today}

\begin{abstract}%

Quantum batteries are quantum systems that potentially serve as fuel for other quantum devices by receiving and supplying energy. In this work, we propose a theoretical scheme to enhance ergotropy and charging power of quantum Battery in the system under consideration through squeezed light. Moreover, The system comprises a coherently driven charger cavity, a strongly dissipative auxiliary cavity acting as a mediator, and an energy--storage battery cavity. Quantum nonreciprocity and directional energy flow are established via destructive quantum interference between the direct coupling path and the loss--engineered indirect path. A degenerate optical parametric amplifier integrated into the charger cavity acts as an active, phase--sensitive energy source that injects two--photon quantum excitations and anomalous correlations. Using a strictly open--quantum--system framework derived directly from the Lindblad master equation, we obtain the full expectation--value dynamics and stationary states. We show that active parametric driving drastically increases the stored energy and peak charging power while maintaining nonreciprocal energy trapping in the battery. We achieve a multi-order-of-magnitude boost in the battery's stored energy, charging power, and ergotropy. We also analyze the effects of pure dephasing on the charging power, ergotropy, and extractable work quality.. Although two--photon pumping introduces photon--number fluctuations that slightly lower the fractional ergotropy ratio ${\cal W}_{\rm B}/{\cal E}_{\rm B}$, the absolute extractable work remains substantially elevated compared to unassisted configurations.

\end{abstract}

\maketitle

\section{Introduction} %

Continuous--variable (CV) quantum systems offer a foundational framework for quantum optics and information processing using continuous field quadratures. Gaussian states--such as coherent, thermal, and squeezed states--are fully described by their first moments and covariance matrices, enabling straightforward modeling of bosonic dynamics under linear and quadratic operations \cite{Weedbrook2012,Adesso2014}. Implemented via accessible optical, microwave, cavity QED, and cavity optomechanical platforms \cite{Vitali07,Mauro07,Groblacher09,Aspelmeyer2014}, Gaussian CV systems provide direct access to symplectic eigenvalues, quantum correlations, and phase--space thermodynamics, serving as a powerful bridge between quantum communication, sensing, and nonequilibrium thermodynamics \cite{Paris09,Weedbrook2012,Adesso2014,Mari15, Amghar24,Harraf2025, Harraf26}.\\

Degenerate optical parametric amplifier (OPA) serves as a powerful Gaussian resource by driving nonlinear interactions that generate phase-sensitive amplification and quadrature squeezing. Moreover, crucial for quantum-limited measurement and nonclassical state generation \cite{Wu1986,Malnou2018}, the pump phase acts as a key control parameter to orient squeezed quadratures and actively tune system energy and fluctuation dynamics. Experimentally, these parametric interactions are readily implemented via optical nonlinearities or Josephson elements in cavity and circuit-QED architectures \cite{Malnou2018,Agarwal2016}. At the same time, nonreciprocity has emerged as an important ingredient for controlling the directional propagation of quantum signals and energy in photonic and cavity systems \cite{Barzanjeh25}. Nonreciprocal interactions can be generated using interference, synthetic gauge fields, reservoir engineering, and engineered dissipation \cite{Metelmann2015}. In particular, reservoir engineering has demonstrated that coherent and dissipative interactions can be combined to produce directional cavity dynamics without requiring intrinsically nonreciprocal microscopic couplings \cite{Metelmann2015}. A wide range of physical strategies enables directionality and nonreciprocal transport, spanning reservoir engineering approaches \cite{Ahmadi2024, Metelmann2015, Fang2017, Zhang2021, Poyatos1996, Begg2024}, magneto-optic effects \cite{Bi2011}, Laser frequency stabilization in a nonreciprocal resonator \cite{Fan2012}, loss--induced quantum nonreciprocity \cite{Li2024}, nonreciprocal quantum synchronization \cite{Lai2025} and nonlinear optical processes \cite{Khanikaev2015}. Directional operation can also be established by leveraging quantum squeezing effects, non-Hermitian dynamics, and topological or chiral photonic platforms \cite{Wang2025}. More recently, loss engineering has provided a particularly appealing route toward nonreciprocity, where a lossy intermediate mode modifies the phase of an indirect transmission pathway and enables destructive interference between direct and indirect channels \cite{Huang2023}. This mechanism is especially attractive for cavity-based quantum technologies because dissipation, rather than being treated solely as a deleterious effect, becomes an active resource for controlling the direction of quantum transport.\\

These developments naturally connect continuous-variable quantum optics with quantum thermodynamics, where the central questions concern the storage, transfer, transformation, and extraction of energy at the quantum level. Quantum batteries \cite{Alicki2013, Quach2022, Yang2024, Song2024, Ahmadi2024, Hadipour2025, Zafar2026, Lai2024,Campaioli2018,Andolina2019,Santos2019} provide a particularly useful framework in this context because they allow the energetic role of coherence, quantum correlations, collective interactions, and nonequilibrium driving to be investigated quantitatively. The performance of a quantum battery cannot, however, be characterized solely by its average stored energy. Recent years have seen substantial progress toward these targets, highlighted by notable gains in ergotropy \cite{Quach2022, Yang2024, Malavazi2025}, accelerated charging rates, and near--lossless energetic dynamics. Numerous quantum battery configurations have been studied, from many-body systems \cite{Le2018} and Dicke configurations \cite{Ferraro2018} to platforms built on ultracold atoms \cite{RojoFrancas2024}, Sachdev-Ye-Kitaev models, and Tavis--Cummings cavities \cite{Yang2024}. Unlocking deeper control over these setups has relied on diverse strategies, including intelligent optimization via reinforcement learning or optimal control theory \cite{Mazzoncini2023}, alongside physical mechanisms such as non--Markovian memory effects, and wireless charging schemes \cite{Song2024}. Besides, the signature quantum advantage a collective, quadratic enhancement in charging power has been rigorously established theoretically and confirmed experimentally in superconducting qubit arrays and quantum optical setups \cite{Ferraro2018, Hadipour2025}. The distinction between internal energy and extractable work is fundamental, with the latter being quantified by the ergotropy, namely the maximum work that can be extracted through cyclic unitary transformations \cite{Allahverdyan2004}. Previous studies have demonstrated that quantum resources can enhance the charging power of batteries and that charger-mediated energy transfer provides a natural framework for separating the charger from the energy-storage system. More recently, continuous--variable quantum batteries have attracted increasing attention because bosonic modes provide experimentally accessible systems in which displacement, squeezing, nonlinear couplings, and dissipation can be controlled. For example, multimode continuous--variable batteries have been shown to exhibit nontrivial charging properties and scaling of energy fluctuations.
Squeezing has also been identified as a mechanism capable of strongly enhancing the stored energy and ergotropy of bosonic quantum batteries and of mitigating the detrimental role of dissipation. Furthermore, work extraction presents an interesting thermodynamic constraint, since not every energetic contribution of a Gaussian state is necessarily convertible into useful work under restricted Gaussian operations. Recently, Gaussian quantum battery is of paramount importance in quantum information processing. These results indicate that continuous--variable quantum batteries provide a natural arena in which energy, fluctuations, squeezing, coherence, dissipation, and extractable work can be studied within a single physical platform. Nevertheless, previous studies have generally considered either continuous-variable charging enhanced by squeezing or nonreciprocal energy--transfer architectures, whereas the simultaneous combination of loss--engineered directionality and active parametric amplification has received considerably less attention.\\

Recent studies have established several strategies for enhancing nonreciprocal quantum-battery performance, including reservoir-engineered directional charging \cite{Ahmadi2024} and loss-induced nonreciprocity mediated by an auxiliary cavity \cite{Zafar2026}. More recently, quantum battery enhancement via degenerate optical parametric amplifier and common reservoirs \cite{Yang2026}. In contrast to these approaches, the present work combines loss-engineered nonreciprocal energy transfer with a degenerate OPA integrated into the charger, thereby introducing a phase-sensitive source of energy and anomalous correlations. Starting directly from the Lindblad master equation \cite{Lindblad1976,Walls2008}, we investigate how the OPA gain and pump phase influence the stored energy, charging power, and ergotropy, while accounting for the fluctuations induced by parametric driving. This framework provides a route to controlling both the direction of energy transfer and the energetic resources supplied to a bosonic quantum battery.\\

In this work, we introduce a squeezed--light--enhanced, nonreciprocal quantum battery consisting of three bosonic cavity modes: a coherently driven charger, a strongly dissipative auxiliary mediator, and a battery mode. The auxiliary cavity is deliberately subjected to strong dissipation, which modifies the interference between a direct charger--battery pathway and an indirect pathway mediated by the lossy auxiliary mode, thereby generating directional energy transfer. A degenerate optical parametric amplifier is incorporated into the charger and provides an additional phase-sensitive source of two--photon excitations and anomalous correlations. The resulting architecture therefore separates two physical functions: engineered dissipation controls the directionality of energy transfer, whereas parametric amplification controls the amount of energy supplied to the charging network. Starting directly from the Lindblad master equation, we derive the equations governing the first moments, populations, normal correlations, and anomalous correlations of the three--mode system.  We then investigate the effects of the OPA gain, pump phase, coupling phase, and auxiliary dissipation on the stationary battery energy and the directionality of energy transfer. Our results show that OPA strongly enhances the transient charging power, can substantially increase the energy stored in the battery while preserving the loss--engineered directional transfer mechanism. We further characterize the battery through its ergotropy and show that the absolute extractable work can remain strongly enhanced even when the fractional ratio between ergotropy and stored energy is reduced by the additional fluctuations and mixedness generated by the active parametric drive. Furthermore, we also consider the case in which the coherent drive is absent, showing that parametric excitation itself can supply the energy required for charging. The proposed architecture can be implemented using experimentally accessible cavity--QED, circuit--QED, or integrated nonlinear photonic platforms, thereby providing a possible route toward fast and directional bosonic energy-storage devices.\\

The paper of this paper is organized as follows. Section~\ref{sec1} introduces the three-mode cavity model and the corresponding Lindblad master equation, and derives the equations governing the first and second moments. Section~\ref{sec2} presents the numerical results, including the transient and stationary charging dynamics, the dependence on the OPA gain and coupling parameters, the role of the auxiliary dissipation, and the charging power and ergotropy. The effects of pure dephasing on the charging performance and extractable work are also analyzed in this section.  Concluding results and remarks close this paper.

\begin{figure}
\includegraphics[width=16cm]{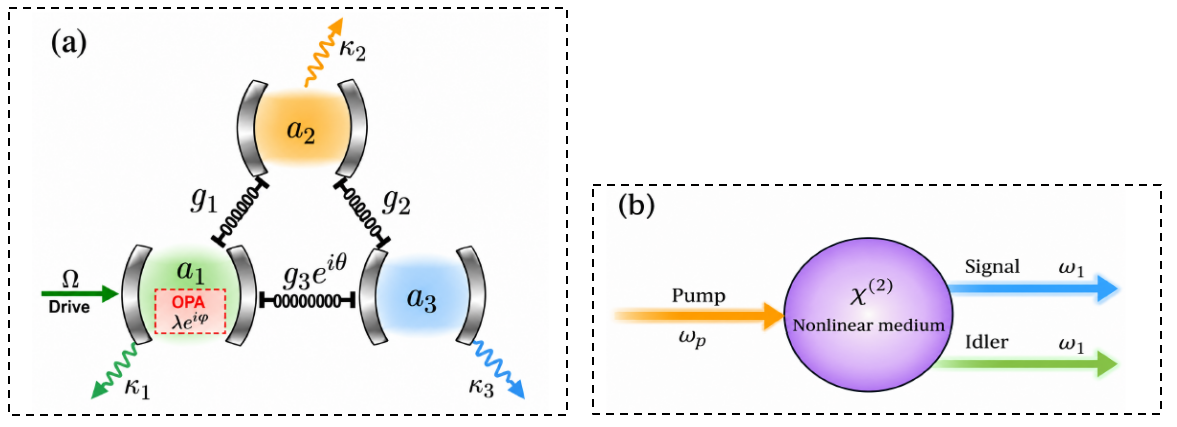}
\caption{(a) Schematic representation of the system battery under consideration. (b)  The system utilizes a degenerate OPA process where, mediated by second-order nonlinearity $\chi^{(2)}$, a single pump photon of frequency $\omega_{\rm p}$ is converted into two cavity photons of frequency $\omega_1$ (signal and Idler with same frequency $\omega_1$). This process is governed by the resonance relation $\omega_{\rm p}=2\omega_1$.}
\label{fig1}
\end{figure}

\section{Model and Lindblad dynamics} \label{sec1}

We now introduce the three--mode bosonic architecture underlying the proposed quantum battery and formulate its dynamics within a Lindblad open--system framework. The model separates the roles of coherent charging, dissipative mediation, and energy storage, while the OPA provides an additional active source of bosonic excitations. Moreover, the system does not depend on common--reservoir dissipation to mediate interactions. Instead, the cavities are fully coherent and coupled directly, while decay occurs strictly via local dissipation.

The physical setup of the system composed of three cavities is schematically illustrated in Fig.~\ref{fig1}(a). The three cavities modes are described by the bosonic annihilation operators ${\rm a}_1$, ${\rm a}_2$ and ${\rm a}_3$, with corresponding creation operators ${\rm a}_1^\dagger$, ${\rm a}_2^\dagger$ and ${\rm a}_3^\dagger$, obeying the standard commutation relations $[{o}, {o}^\dagger] = 1$ for ${o} \in \{{\rm a}_1, {\rm a}_2, {\rm a}_3\}$. Mode~$\rm a_1$ acts as the charger, mode~$\rm a_2$ serves as a highly dissipative auxiliary mediator ($\kappa_2 \gg \kappa_{1,3}$), and mode~$\rm a_3$ constitutes the energy storage battery. The charger is coherently driven by an external laser with amplitude $\Omega$ and frequency $\omega_d$, and contains a second-order nonlinear $\chi^{(2)}$ medium providing degenerate optical parametric amplification with gain $\lambda$ and pump phase $\phi$ (e.g., lithium niobate or aluminum nitride \cite{Wolf17,Szabados20,Wang20}). Coherent intermode couplings $\rm g_{1}$ and $\rm g_{2}$ connect the modes sequentially, while a direct phase-dependent coupling $\rm g_{3}e^{i\theta}$ links the charger directly to the battery. The essential physical mechanism driving directional energy transfer is the quantum interference between two distinct energy-transfer pathways: the direct channel $\rm a_1 \to \rm a_3$ and the indirect channel $\rm a_1 \to \rm a_2 \to \rm a_3$. The auxiliary cavity's strong decay rate $\kappa_2$ introduces a dissipation-induced phase shift. Concurrently, the OPA acts as an active quantum resource, continuously generating photon pairs via parametric down-conversion ($\omega_{\rm p} = 2\omega_1$) as shown in Fig.~\ref{fig1}(b), which injects phase-sensitive squeezing into the charger. This dual framework separates the system functionality into two distinct mechanisms: loss engineering establishes directional nonreciprocal energy flow from $\rm a_1$ to $\rm a_3$, while active parametric driving enhances the total excitation level and charging performance. In a frame rotating at the frequency $\omega_d$ of the coherent drive, the Hamiltonian of system is written as
\begin{align}
{\cal H}={}&
\Delta_{1} {\rm a_1}^\dagger {\rm a_1} + \Delta_2 {\rm a_2}^\dagger {\rm a_2} + \Delta_3 {\rm a_3}^\dagger {\rm a_3}
\nonumber\\
&+
{\rm g}_1(\rm a_1 a_2^\dagger+\rm a_1^\dagger \rm a_2)
+
{\rm g}_2(\rm a_3 a_2^\dagger+\rm a_3^\dagger \rm a_2)
\nonumber\\
&+
{\rm g}_3
\left(
e^{\I\theta}\rm a_1 a_3^\dagger+
e^{-\I\theta}\rm a_1^\dagger \rm a_3
\right) + \frac{\I\lambda}{2}
\left(
e^{-\I\phi}\rm a_1^{\dagger 2}
-
e^{\I\phi}\rm a_1^2
\right)
\nonumber\\
&+
\Omega(\rm a_1 + \rm a_1^\dagger),
\label{eq:H0}
\end{align}
where $\Delta_j=\omega_j-\omega_d, \quad j=1,2,3$, and $\lambda$ is the parametric gain and $\phi$ is the parametric phase. The density operator obeys the local Lindblad master equation
\begin{equation}
\dot{\varrho} = -\I[{\cal H},\varrho] + \kappa_{\rm a_1}\D[\rm a^\dag_1]\varrho+ \kappa_{a_2}\D[\rm a^2\dag_2]\varrho+ \kappa_{a_3}\D[\rm a^\dag_3] \varrho,
\label{eq:LME}
\end{equation}
with $\D[o]\varrho = o\varrho o^\dagger - \frac12o^\dagger o\varrho - \frac12\varrho o^\dagger o$ and $\varrho(t)$ is the density operator of the system and $\kappa_j (j=1,2)$ are the dissipation constants representing the energy dissipation from the each cavity to the environment. The charger is in contact with a thermal bath as an environment that induces transfer of population between them~.  We consider a low-temperature setup of the bath that allows us to neglect the population transfer from the environment to the charger or thermal excitation. We also focus on the coherent driving for the excitation of the charger, since the thermal excitation does not contribute to extractable energy from the battery~. By solving the master equation, the expectation value of the operator can be calculated as $\langle O\rangle(t)=\textrm{Tr}[O\varrho(t)]$, with $\rm{Tr}[ABC]=\rm{Tr}[CAB]=\rm{Tr}[BCA]$. After charging, the battery is disconnected from the charger. We want the battery to store maximum energy within minimal charging time. Thus, the appropriate quantities to characterize the performance of the battery are the maximum stored energy and the time it takes to reach this maximum energy is referred to as optimal charging time~. The average energy stored by the battery is expressed as follows
\begin{align}
    {\cal E}_{\rm B}(t) = \hbar\omega_3 \langle \rm a_3^\dag \rm a_3\rangle(t).
    \label{eq:eb}
\end{align}
The optimal charging time ($t_\textrm{max}$) is defined as the time to reach the maximum ${\cal E}_{\rm B}$, i.e., $ {\cal E}_{\rm B}(t_\textrm{max})= {\cal E}_\textrm{max}$~. The charger energy is similarly ${\cal E}_{\mathrm ch}(t) = \hbar\omega_1 \langle \rm a_1^\dag \rm a_1\rangle(t)$. The quantity measures how much energy is present in the battery relative to the charger is given by
\begin{equation}
\xi_{13}(t) = \frac{{\cal E}_{\mathrm B}(t)}{{\cal E}_{\mathrm ch}(t)}.
\label{eq:gain}
\end{equation}
Consequently, a large value of $\xi_{13}(t)$ does not determine the fraction of the total externally supplied energy that reaches the battery. Regions with $\xi_{13}(t)>1$ indicate that the battery occupation exceeds the charger occupation in the stationary state; regions with $\xi_{13}(t)<1$ indicate the opposite. Moreover, the instantaneous charging power is
\begin{equation}
{\cal P}_{\mathrm B}(t) = \frac{\dd {\cal E}_{\mathrm B}}{\dd t}.
%
\end{equation}
It should be noted that not all of the transferred energy can be used for work. The maximal amount of energy that can be extracted for work is known as ergotropy~\cite{Allahverdyan2004}. The ergotropy can be obtained by considering the energy of the passive state of the battery, from which no work can be extracted. It is useful to distinguish stored energy from extractable work. For a battery state $\rho_{\mathrm B}$ with eigenvalues $r_k$ arranged in decreasing order, i.e., $r_0\geq r_1\geq ...$, and battery-energy eigenstates $\ket{\epsilon_k}$ arranged in increasing energy, i.e., $\epsilon_0\leq \epsilon_1\leq ...$, the passive state is $\rho_{\mathrm B}^{\mathrm{pas}} = \sum_k r_k\ket{\epsilon_k}\bra{\epsilon_k}.$
The ergotropy is given by
\begin{equation}
\mathcal W_{\mathrm B} = \Tr(\varrho_{\mathrm B}{\cal H}_{\mathrm B}) - \Tr(\varrho_{\mathrm B}^{\mathrm{pas}}{\cal H}_{\mathrm B}).
\label{eq:ergotropy}
\end{equation}
For an arbitrary system operator $O$, using the equation (Eq.~\eqref{eq:LME}), the equations of motion for $\alpha=\langle \rm a_1\rangle,\quad \beta=\langle \rm a_2\rangle,\quad \gamma=\langle \rm a_3\rangle.$
\begin{align}
\dot{\alpha}
={}&
-\left(
\I\Delta_1+\frac{\kappa_1}{2}
\right)\alpha
-\I {\rm g}_1\beta
-\I {\rm g}_{2}e^{-\I\theta}\gamma
-\I\Omega
+\lambda e^{-\I\phi}\alpha^*,
\label{eq:alpha}
\\
\dot{\beta}
={}&
-\left(
\I\Delta_2+\frac{\kappa_2}{2}
\right)\beta
-\I {\rm g}_{1}\alpha
-\I {\rm g}_{2}\gamma,
\label{eq:beta}
\\
\dot{\gamma}
={}&
-\left(
\I\Delta_3+\frac{\kappa_3}{2}
\right)\gamma
-\I {\rm g}_{3}e^{\I\theta}\alpha
-\I {\rm g}_{2}\beta.
\label{eq:gamma}
\end{align}
and $n_1=\langle {\rm a}_1^\dagger \rm a_1\rangle, n_2=\langle \rm a_2^\dagger a_2\rangle, n_3=\langle \rm a_3^\dagger a_3\rangle$. Application of Eq.~\eqref{eq:generalO2} gives
\begin{align}
\dot n_1={}& -\kappa_1 n_1-2{\rm g}_1\Im\langle {\rm a_1}{\rm a_2}^\dagger\rangle-2{\rm g}_{3}\Im\left[e^{-\I\theta}\langle {\rm a_1}{\rm a_3}^\dagger\rangle\right]-2\Omega\Im\alpha+2\lambda\Re\left[e^{-\I\phi}\langle {\rm a_1}^{\dagger2}\rangle\right],
\label{eq:na1}
\\
\dot n_2={}&-\kappa_2 n_2+2{\rm g}_{1}\Im\langle {\rm a_1}{\rm a_2}^\dagger\rangle
+2{\rm g}_{2}\Im\langle {\rm a_3}{\rm a_2}^\dagger\rangle,
\label{eq:na2}
\\
\dot n_3={}&-\kappa_3 n_3-2{\rm g}_{2}\Im\langle {\rm a_3}{\rm a_2}^\dagger\rangle+2{\rm g}_{3}\Im\left[e^{\I\theta}\langle {\rm a_1}{\rm a_3}^\dagger\rangle\right].
\label{eq:na3}
\end{align}

The OPA does not appear explicitly in $\dot n_3$; its influence reaches the battery through the correlations generated in the charger. We have also 
$P=\langle {\rm a_1}{\rm a}_2^\dagger\rangle, Q=\langle {\rm a_1}{\rm a}_3^\dagger\rangle, R=\langle {\rm a_3}{\rm a}_2^\dagger\rangle$.
Their equations are
\begin{align}
\dot P
={}&
-\left[
\I(\Delta_1-\Delta_2)
+\frac{\kappa_1+\kappa_2}{2}
\right]P +
\I g_{1}(n_1-n_2)
+\I g_{2}Q
-\I g_{3}e^{-\I\theta}R
-\I\Omega\beta^*
+\lambda e^{-\I\phi}X^*,
\label{eq:P}
\\
\dot Q
={}&
-\left[
\I(\Delta_1-\Delta_3)
+\frac{\kappa_1+\kappa_3}{2}
\right]Q+
\I g_{2}P
-\I g_{1}R^*
+\I g_{3}e^{-\I\theta}(n_1-n_3)
-\I\Omega\gamma^*
+\lambda e^{-\I\phi}Y^*,
\label{eq:Q}
\\
\dot R
={}&
-\left[
\I(\Delta_3-\Delta_2)
+\frac{\kappa_3+\kappa_2}{2}
\right]R+
\I g_{2}(n_3-n_2)
-\I g_{3}e^{\I\theta}P
+\I g_{1}Y^*.
\label{eq:R}
\end{align}
Here we have introduced the anomalous correlations $X=\langle {\rm a_1}{\rm a_2}\rangle, Y=\langle {\rm a_1}{\rm a_3}\rangle.$
The OPA creates pair correlations even if the initial state is the vacuum. We define $
A=\langle {\rm a}_1^2\rangle, B=\langle {\rm a}_2^2\rangle, C=\langle {\rm a}_3^2\rangle$, and $Z=\langle {\rm a_2}{\rm a_3}\rangle$. The corresponding equations are
\begin{align}
\dot A
={}&
-(2\I\Delta_1+\kappa_1)A
-2\I {\rm g}_{1}X
-2\I {\rm g}_{3}e^{-\I\theta}Y-2\I\Omega\alpha
+2\lambda e^{-\I\phi}(n_1+1),
\label{eq:A}
\\
\dot B
={}&
-(2\I\Delta_2+\kappa_2)B
-2\I {\rm g}_{1}X
-2\I {\rm g}_{2}Z,
\label{eq:B}
\\
\dot C
={}&
-(2\I\Delta_3+\kappa_3)C
-2\I {\rm g}_{3}e^{\I\theta}Y
-2\I {\rm g}_{2}Z,
\label{eq:C}
\\
\dot X
={}&
-\left[
\I(\Delta_1+\Delta_2)
+\frac{\kappa_1+\kappa_2}{2}
\right]X
-\I {\rm g}_{1}(A+B)
-\I {\rm g}_{3}e^{-\I\theta}Z
-\I {\rm g}_{2}Y
-\I\Omega\beta
+\lambda e^{-\I\phi}P^*,
\label{eq:X}
\\
\dot Y
={}&
-\left[
\I(\Delta_1+\Delta_3)
+\frac{\kappa_1+\kappa_3}{2}
\right]Y
-\I {\rm g}_{1}Z
-\I {\rm g}_{3}
\left(
e^{-\I\theta}C+e^{\I\theta}A
\right)
\nonumber\\
&-\I {\rm g}_{2}X
-\I\Omega\gamma
+\lambda e^{-\I\phi}Q^*,
\label{eq:Y}
\\
\dot Z
={}&
-\left[
\I(\Delta_2+\Delta_3)
+\frac{\kappa_2+\kappa_3}{2}
\right]Z
-\I {\rm g}_{1}Y
-\I {\rm g}_{2}(B+C)
-\I {\rm g}_{3}e^{\I\theta}X.
\label{eq:Z}
\end{align}

\section{Numerical results and discussions}  \label{sec2}

Having established the dynamical equations of the three-mode system, we next investigate its charging performance and the mechanisms responsible for directional energy accumulation. We focus on the roles of parametric amplification, coherent coupling, dissipation, and phase control in determining the stored energy and extractable work.

\begin{table}[t]
\caption{Parameter configuration for the three--mode system \cite{Li2024,Zafar2026}.}
\label{table1}
\begin{ruledtabular}
\begin{tabular}{lll}
Parameter & Symbol  & Value \\
\hline
Detunings & $\Delta_1,\Delta_2,\Delta_3$ & $\Delta_1=\Delta_2=\Delta_3=0$\\
Inter-cavity couplings & $g_1,g_2,g_3$ & $g_1=g_2=g_3=g=\sqrt2\,\kappa$\\
Decay rates & $\kappa_1,\kappa_2,\kappa_3$ & $\kappa_1=\kappa_3=\kappa,\ \kappa_2=10\kappa$\\
Coupling phase & $\theta$ & $-\pi/2$\\
Coherent charger amplitude & $\Omega$ & $0.5\kappa$\\
OPA gain and phase & $\lambda,\phi$ & $\lambda=\kappa,\ \phi=0$\\
\end{tabular}
\end{ruledtabular}
\end{table}

\begin{figure}[!htb]
\minipage{0.5\textwidth}
\includegraphics[width=1\linewidth]{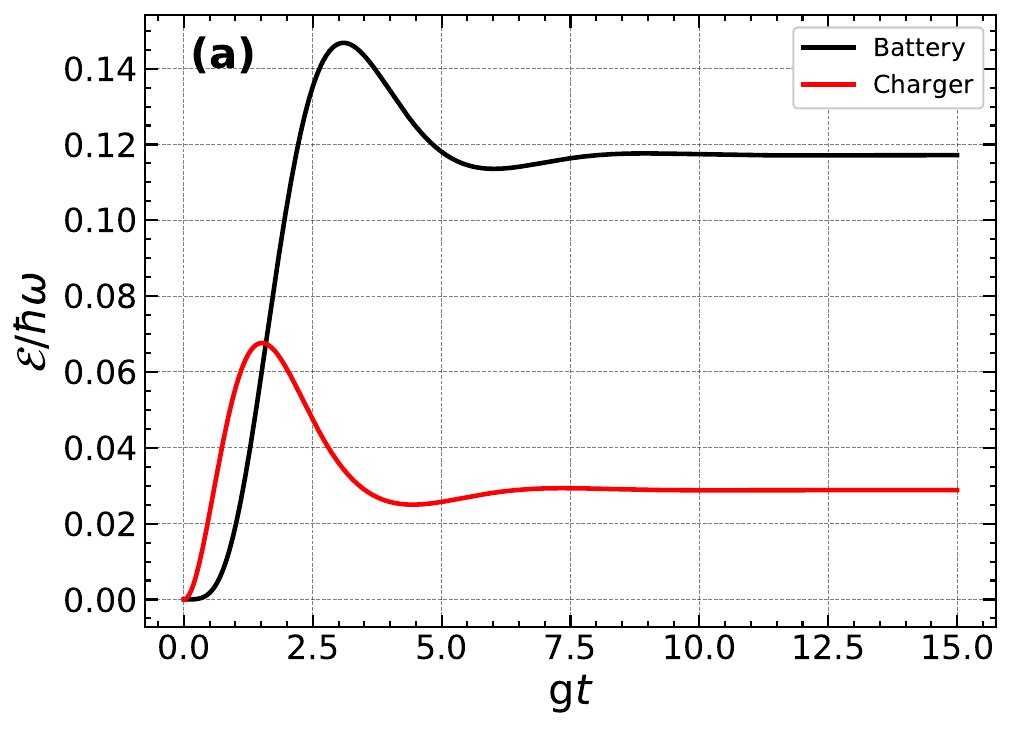}
\endminipage\hfill
\minipage{0.5\textwidth}
\includegraphics[width=1\linewidth]{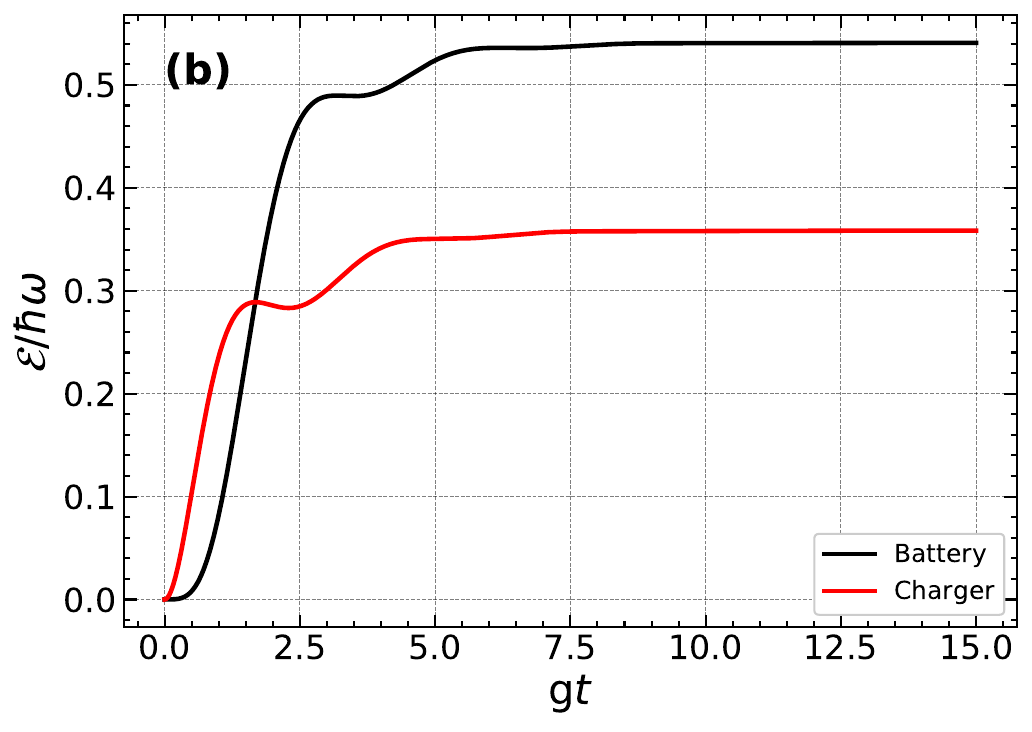}
\endminipage\hfill
\minipage{0.5\textwidth}
\includegraphics[width=1\linewidth]{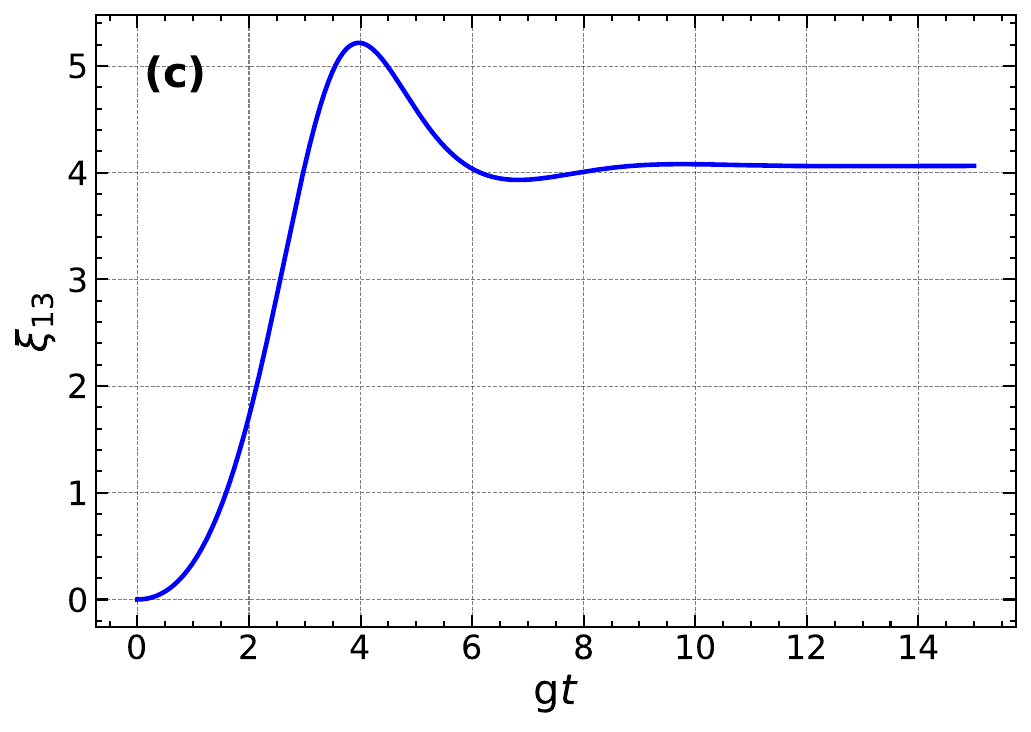}
\endminipage\hfill
\minipage{0.5\textwidth}
\includegraphics[width=1\linewidth]{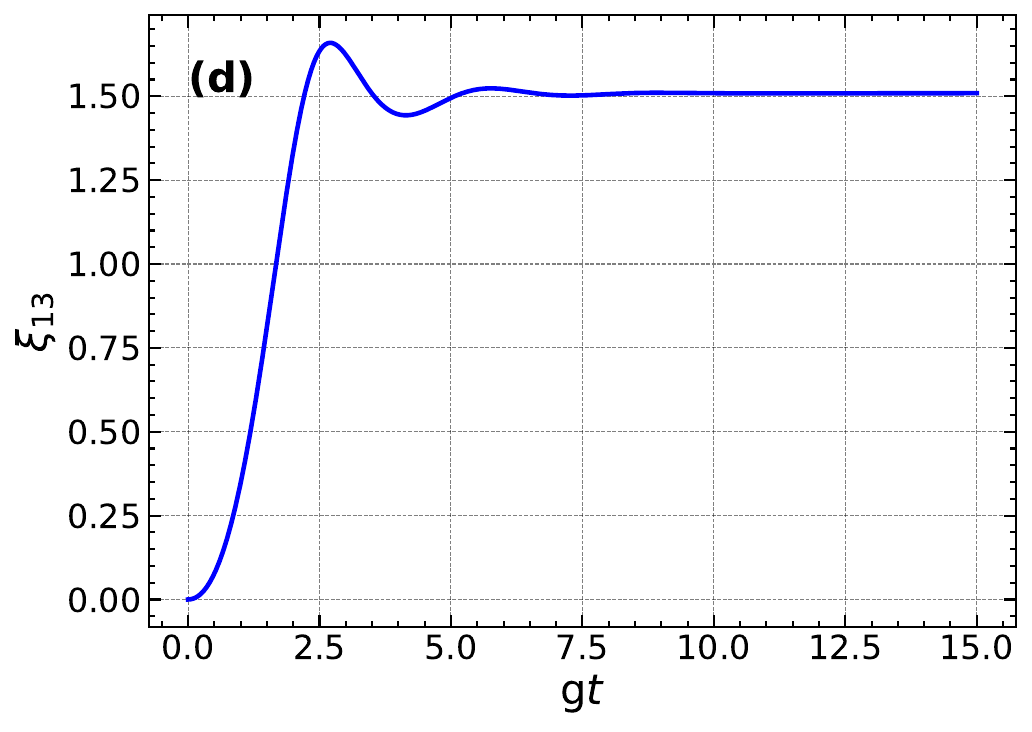}
\endminipage\hfill
\caption{Dynamical evolution of the charging of the quantum battery. (a,b) Stored energies of the charger and battery, ${\cal E}_{\rm ch}$ and ${\cal E}_{\rm B}$, without ($\lambda=0$) (a) and with the OPA ($\lambda/\kappa =1$) (b), respectively. (c,d) Corresponding energy ratios $\xi_{13} = {\cal E}_{\rm B}/{\cal E}_{\rm ch}$. The other parameters are the same as in the table \ref{table1}.} 
\label{fig2}
\end{figure}

Fig.~\ref{fig2} presents the temporal evolution of the stored energies ${\cal E}_{\rm ch}(t) = \hbar\omega_1 \langle {\rm a_1}^\dagger {\rm a_1} \rangle$ and ${\cal E}_{\rm B}(t) = \hbar\omega_2 \langle {\rm a_3}^\dagger {\rm a_3} \rangle$, together with the energy ratio $\xi_{13} = {\cal E}_{\rm B}/{\cal E}_{\rm ch}$, starting from the joint vacuum state. This figure shows that in the absence of parametric driving ($\lambda = 0$), panels (a) and (c) show that the battery population increases monotonically after transient Rabi--like oscillations and eventually exceeds that of the charger, reaching a steady--state ratio $\xi_{13} > 1$. This means that this energy accumulation asymmetry engineered by the lossy auxiliary mode ${\rm a_2}$. However, when the OPA is activated ($\lambda = \kappa$), panels (b) and (d) show a dramatic enhancement in both the absolute stored energy ${\cal E}_B$ and the charger population ${\cal E}_{\rm ch}$. Furthermore, the parametric term introduces an anomalous pumping channel $\propto \frac{\lambda}{2} e^{-i\phi} \langle {\rm a}^{\dagger 2} \rangle$, which continuously feeds energy into the charger mode. Moreover, the underlying loss--engineered transfer remains intact, these additional excitations are efficiently funneled toward the battery. Also, the ratio ${\cal E}_{\rm B}/{\cal E}_{\rm ch}$ retains its elevated magnitude, proving that the nonreciprocal directionality is preserved during active amplification. Besides, the transient oscillations reflect damped coherent energy exchange among the modes before settling into a steady state. Thus, active squeezing significantly expands the battery capacity in comparison with the results reported in Ref.~\cite{Zafar2026}.%

\begin{figure}[!htb]
\minipage{0.5\textwidth}
\includegraphics[width=1\linewidth]{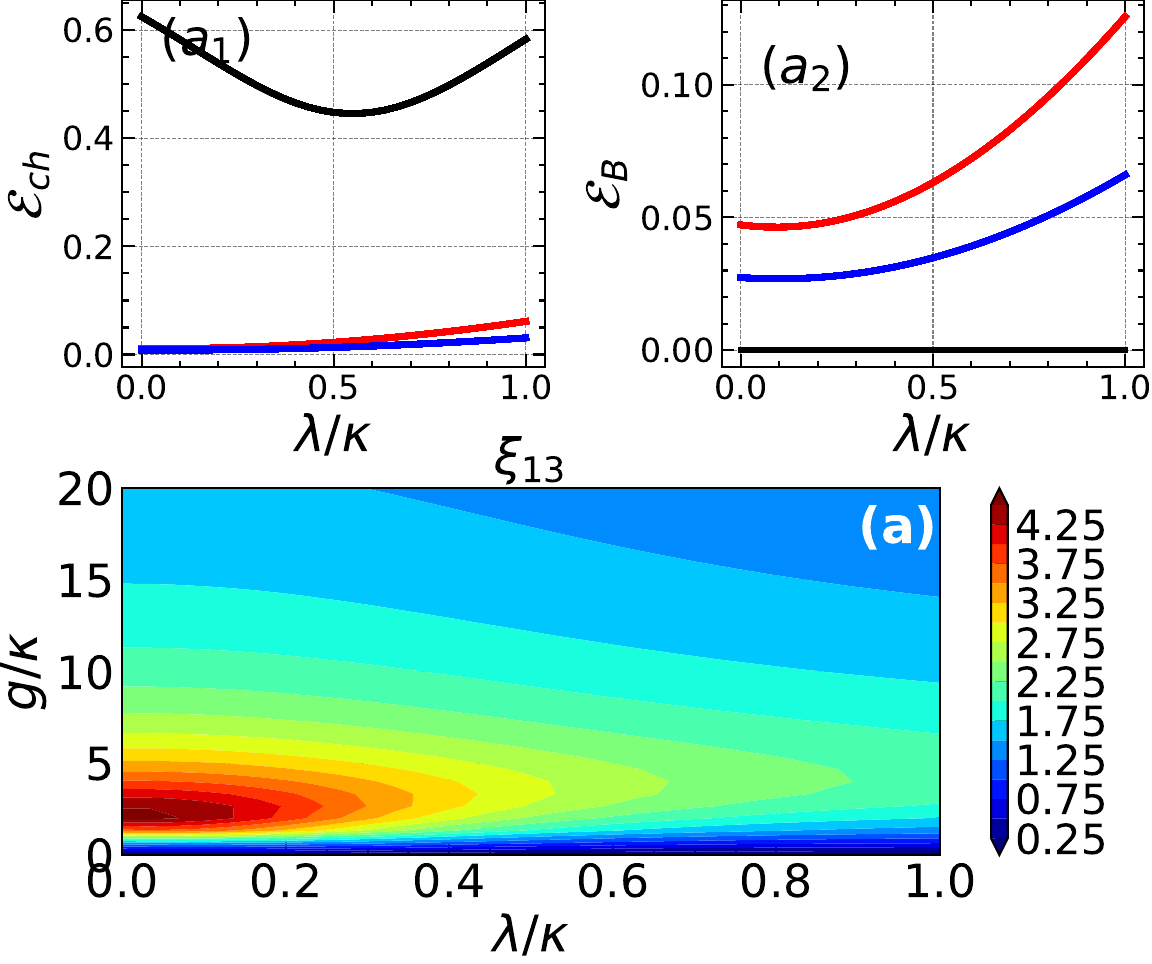}
\endminipage\hfill
\minipage{0.5\textwidth}
\includegraphics[width=1\linewidth]{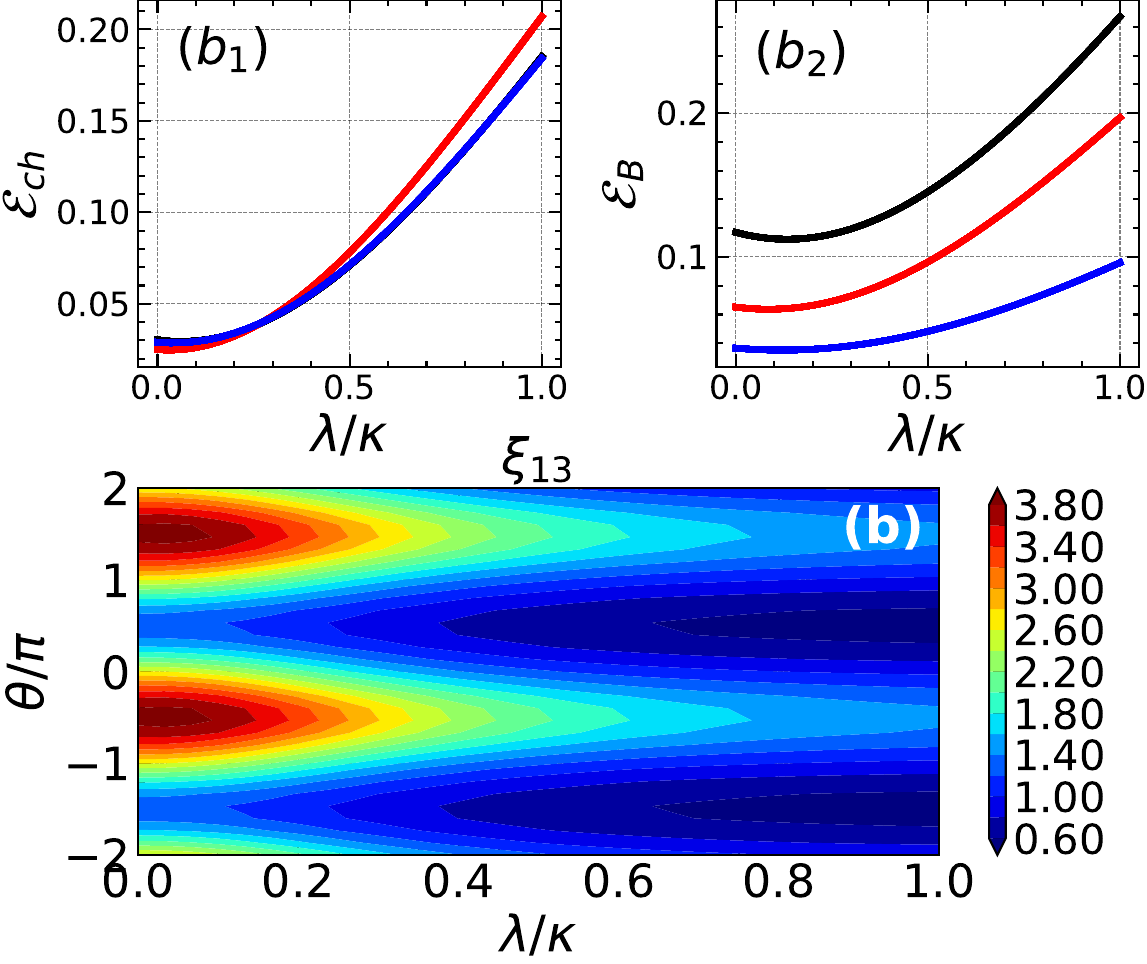}
\endminipage\hfill
\caption{Plot of stationary charging characteristics against the OPA gain $\lambda$
and the coupling parameters $\rm g$. $(\rm a_1,a_2)$ Stationary charger and battery
energies for ${\rm g}=0$ (black curve), $3\kappa$ (red curve), and $5\kappa$ (blue curve). $(b_1,b_2)$ Corresponding stationary charger and battery for $\theta=-\pi/2$ (black curve), $0$ (red curve), and $+\pi/2$ (blue curve). (c) Stationary energy ratio $\xi_{13}$ in the $(\lambda/\kappa,{\rm g}/\kappa)$ plane. (d) Stationary energy ratio in the $(\lambda/\kappa,\theta/\pi)$ plane. The other parameters are the same as in the table \ref{table1}.}
\label{fig3}
\end{figure}

The foremost aim of the investigation depicted in Fig.~\ref{fig3} is to enhance the steady--state stored energies ${\cal E}_{\rm ch}(t)$ and ${\cal E}_{\rm B}(t)$ by tuning the OPA gain $\lambda/\kappa$, the coherent coupling strength $\rm g/\kappa$, and the phase $\theta/\pi$. In Fig.~\ref{fig3} we show how the OPA gain $\lambda$, the intermode coupling strength $\rm g$, and the phase $\theta$ jointly determine the stationary charging performance of the three--mode quantum battery. The discussion distinguishes the stored energy from the energy--transfer ratio $\xi_{13}$, since an increase in the stationary battery population does not necessarily imply an improvement in the relative transfer performance. One can see in subfigures $(\rm  a_1, a_2)$ and $(b_1, b_2)$ display a stationary charger (${\cal E}_{\rm ch}$) and battery (${\cal E}_{\rm B}$) energies as a function of parametric gain $\lambda/\kappa$. 

We first consider the coupling--strength dependence shown in Figs.~\ref{fig3}($\rm a_1,a_2$). In the limiting case ${\rm g}=0$, the intermode transfer channels are absent if all three couplings are set to zero. In this case, the battery remains uncharged, ${\cal E}_{\rm B}$, and the OPA can only excite the charger. This limiting case provides a useful reference for identifying the contribution of coherent intermode coupling. For finite coupling strengths, such as $\rm g=3\kappa$ and $\rm g=5\kappa$, the charger can exchange excitations with the auxiliary mode and the battery. Furthermore, the OPA gain introduces a second control parameter. At $\lambda=0$, the parametric interaction is absent, and the stationary state is sustained by the coherent drive $\Omega$. However, for $\lambda>0$, the OPA provides an additional source of excitations through its pair--creation interaction. Moreover, the stationary charger and battery energies can differ substantially from their values without the OPA. The gain dependence must, however, be interpreted only over the stable interval of $\lambda/\kappa$. Close to the parametric--instability threshold, the linear response can become strongly amplified, and a stationary solution should not be reported beyond the region in which the dynamical stability condition is satisfied.

Panels ($\rm b_1,b_2$) examine the role of the phase $\theta$. For $\theta=0$ provides that the remaining Hamiltonian and dissipative couplings contain no other source of nonreciprocity. By contrast, the choices ($\theta=-\pi/2$) and ($\theta=+\pi/2$) correspond to opposite phase--biased configurations of the direct charger--battery coupling. These phases change the relative phase between the direct pathway and the pathway mediated by the lossy auxiliary cavity, thereby modifying their interference. The cases $\theta=0$ and $\theta=\pm\pi/2$ represent distinct interference conditions between the direct charger--battery pathway and the pathway mediated by the lossy auxiliary mode. Nevertheless, the non--zero phase ($\theta = \pm\pi/2$) alone does not guarantee nonreciprocal energy transfer because single phase parameters can often be eliminated through a unitary transformation (redefining mode operators). Directional energy flow is only genuinely established when fixed coupling and dissipation parameters yield distinctly different forward and reverse system responses.

The lower contour maps (a) and (b) illustrate the steady--state transfer ratio $\xi_{13}$ in the $(\lambda/\kappa, g/\kappa)$ and $(\lambda/\kappa, \theta/\pi)$ parameter planes. Moreover, we observe that $\xi_{13}$ depends non-monotonically on $\rm g/\kappa$: very weak coupling limits energy transfer, whereas excessive coupling ($\rm g \gg \kappa$) overrides the dissipative isolation and restores reciprocal back-action. Panel (b) highlights a pronounced periodicity with respect to the phase $\theta$, showing optimal nonreciprocal energy accumulation at $\theta = -\pi/2 \pm 2n\pi$, with $n\in \mathbb{Z}$. At these phase values, destructive interference completely suppresses backward energy leakage from $\rm a_3$ to $\rm a_1$, while constructive interference maximizes forward transfer $\rm a_1 \to a_3$.%

\begin{figure}[!htb]
\minipage{0.5\textwidth}
\includegraphics[width=1\linewidth]{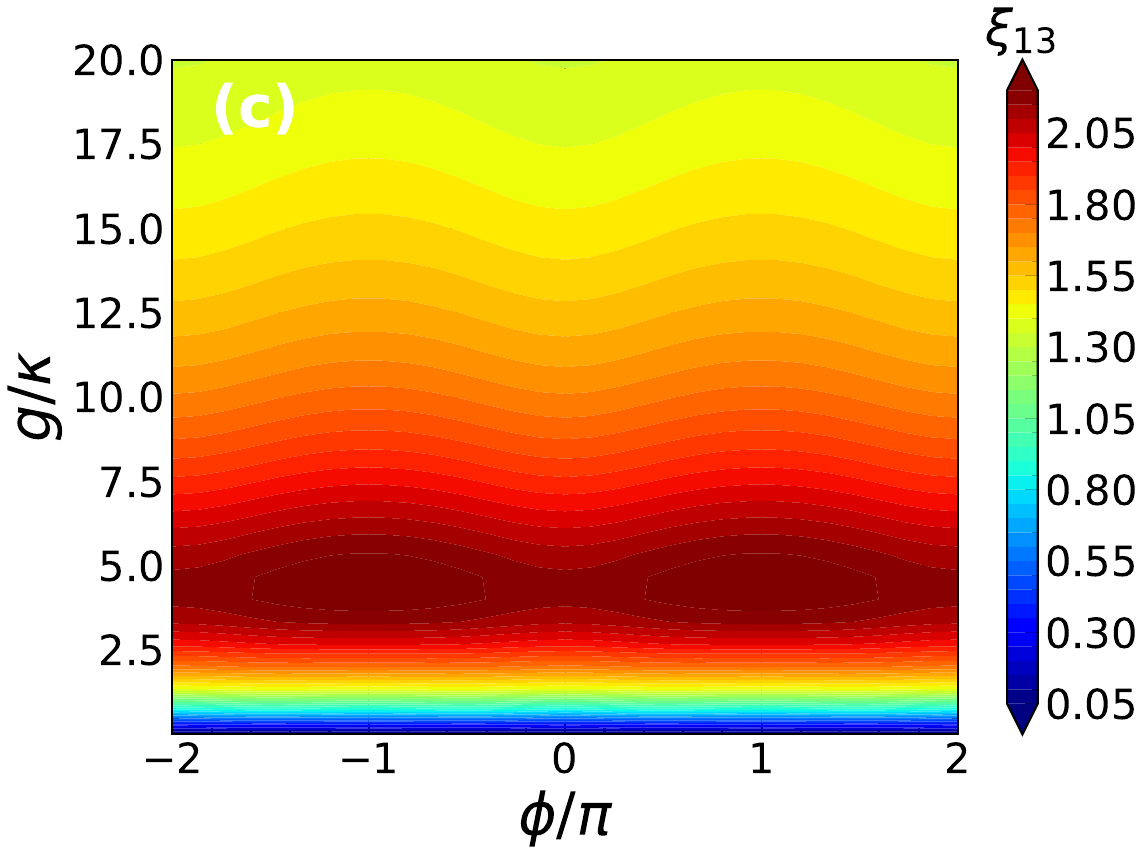}
\endminipage\hfill
\minipage{0.5\textwidth}
\includegraphics[width=1\linewidth]{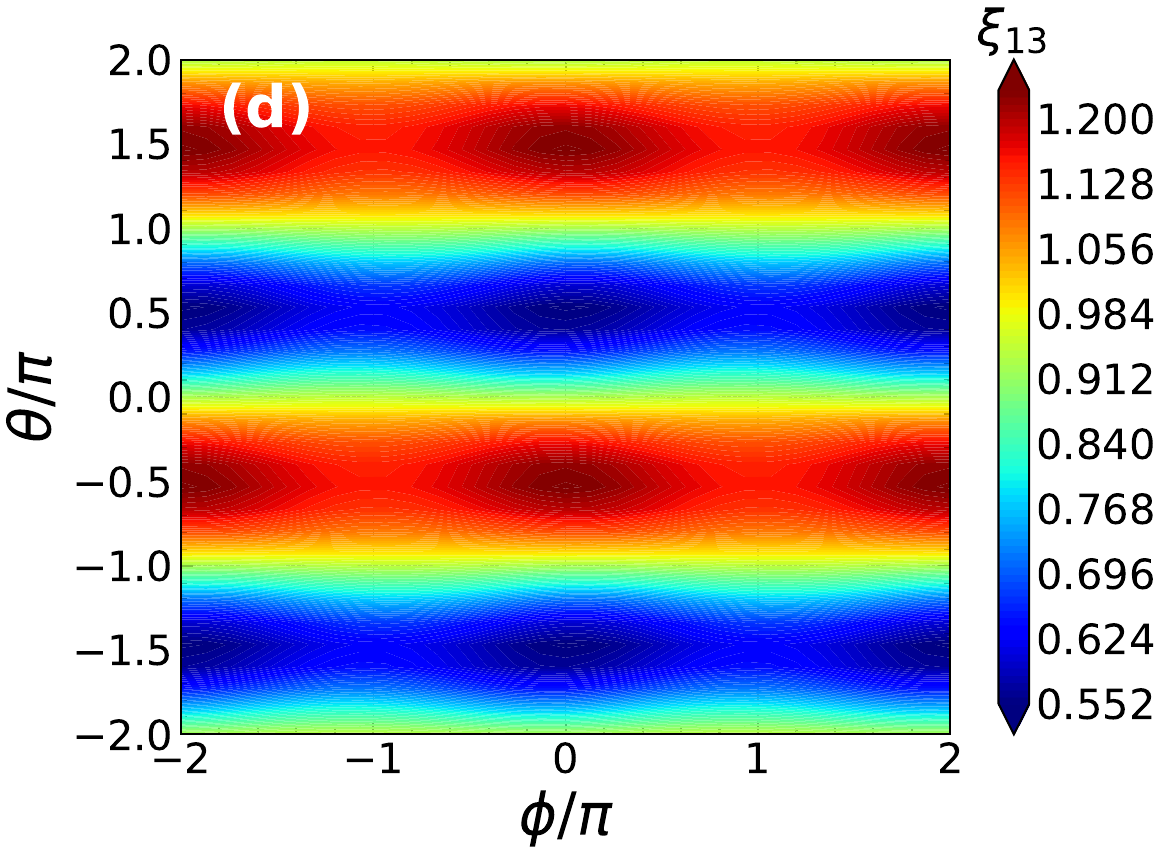}
\endminipage\hfill
\caption{Contour plot of the stationary ratio $\xi_{13}$ as a function the OPA phase $\phi/\pi$ and in panel (c) coupling strength $\rm g/\kappa$ and in panel (d) the phase $\theta/\pi$. The parameters are fixed to the reference values in Table~\ref{table1}.}
\label{fig4}
\end{figure}

To gain a deeper understanding of the phase control and intermode interference mechanisms governing the directional energy dynamics, we present in Figs. \ref{fig4}(c) and \ref{fig4}(d) a contour plot of the steady--state energy ratio $\xi_{13} = \mathcal{E}_{B}/\mathcal{E}_{ch}$.

In Fig. \ref{fig4}(c), the ratio $\xi_{13}$ is mapped across the parameter domain of the normalized OPA pump phase $\alpha/\pi \in [-2, 2]$ and intermode coupling strength $\rm g/\kappa \in [0, 20]$. Remarkably, the contour plot demonstrates that the stationary ratio $\xi_{13}$ remains completely invariant under variations of the OPA phase $\phi/\pi$. Instead, the energy distribution in Fig. \ref{fig4}(c) exhibits a strong non-monotonic functional dependence on the coupling strength $\rm g/\kappa$. In the weak--coupling regime ($\rm g/\kappa < 1$), the energy ratio remains low ($\xi_{13} \ll 1$) due to insufficient intermode excitation transfer. As $\rm g/\kappa$ increases, $\xi_{13}$ rapidly rises to a prominent maximum in the intermediate coupling interval $\rm g/\kappa \in [2, 5]$, where optimal energy accumulation in the battery occurs. Moreover, for strong couplings ($\rm g/\kappa > 10$), the stationary energy ratio gradually degrade and approaches a lower asymptotic threshold. This behavior occurs because excessive coupling overrides the loss--engineered isolation provided by the dissipative auxiliary cavity $\rm a_{2}$, restoring reciprocal back--action between the charger and battery modes.  
 
In sharp contrast, Fig. \ref{fig4}(d) shows the steady-state ratio $\xi_{13}$ in the parameter plane of OPA phase $\phi/\pi \in [-2, 2]$ and coupling phase $\theta/\pi \in [-2, 2]$. One can see that while remaining independent of $\phi/\pi$, $\xi_{13}$ displays pronounced periodic oscillations along the direct coupling phase axis $\theta/\pi$ with a period of $2\pi$. The functional oscillations attain sharp maximum peaks ($\xi_{13} \approx 1.20$) at phase values $\theta/\pi = -\pi/2 + 2n\pi$ ($n \in \mathbb{Z}$) and deep minimum troughs ($\xi_{13} \approx 0.55$) at $\theta/\pi = \pi/2 + 2n\pi$. This oscillatory profile is a direct signature of quantum interference between the direct path $\rm a_1 \to a_3$ (governed by $\rm  g_3 e^{i\theta}$) and the indirect dissipation--engineered path $\rm  a_1 \to a_2 \rightarrow a_3$. At $\theta = -\pi/2 \pm 2n\pi$, destructive quantum interference completely suppresses backward energy leakage from the battery cavity $\rm a_3$ back to the charger $\rm a_1$, trapping transferred excitations inside the battery. However, shifting the phase by $\pi$ to $\theta = \pi/2 \pm 2n\pi$ reverses the interference condition, causing destructive interference along the forward pathway $\rm a_1 \to a_3$ and strongly suppressing the battery charging efficiency.%

\begin{figure}[!htb]
\minipage{0.5\textwidth}
\includegraphics[width=1\linewidth]{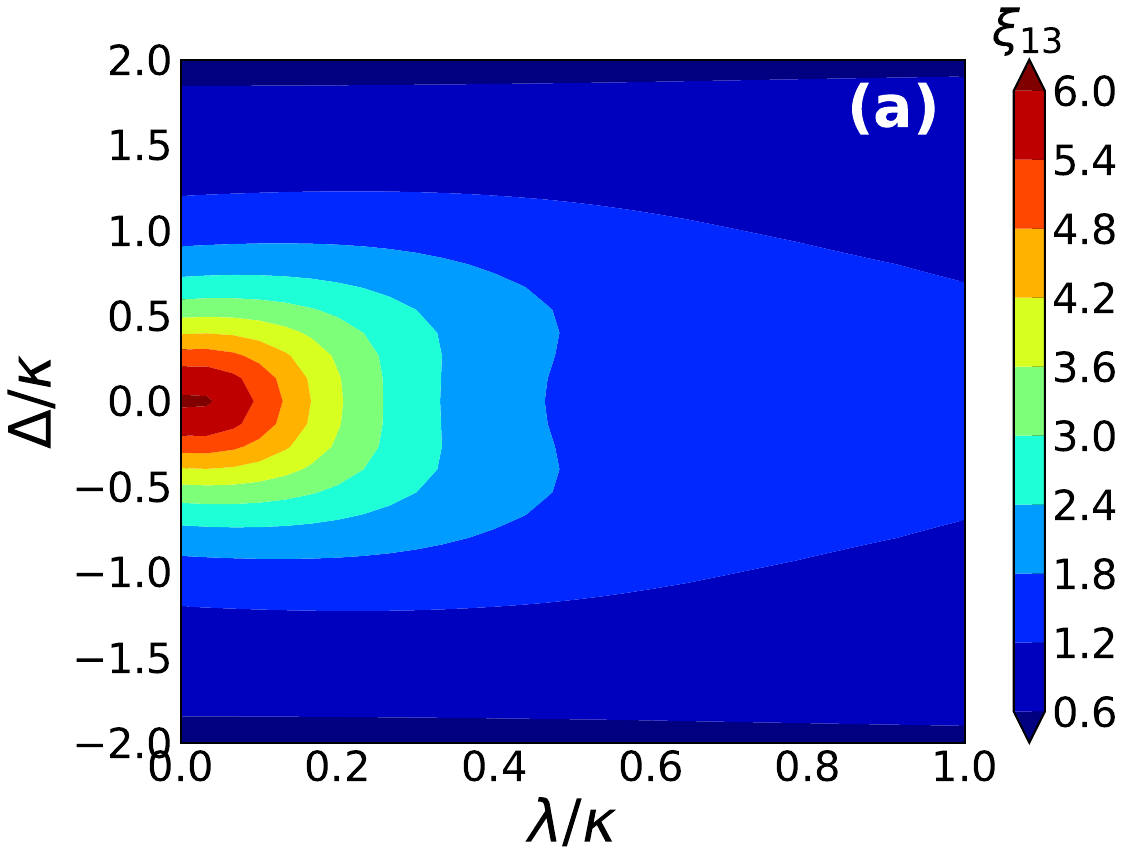}
\endminipage\hfill
\minipage{0.5\textwidth}
\includegraphics[width=1\linewidth]{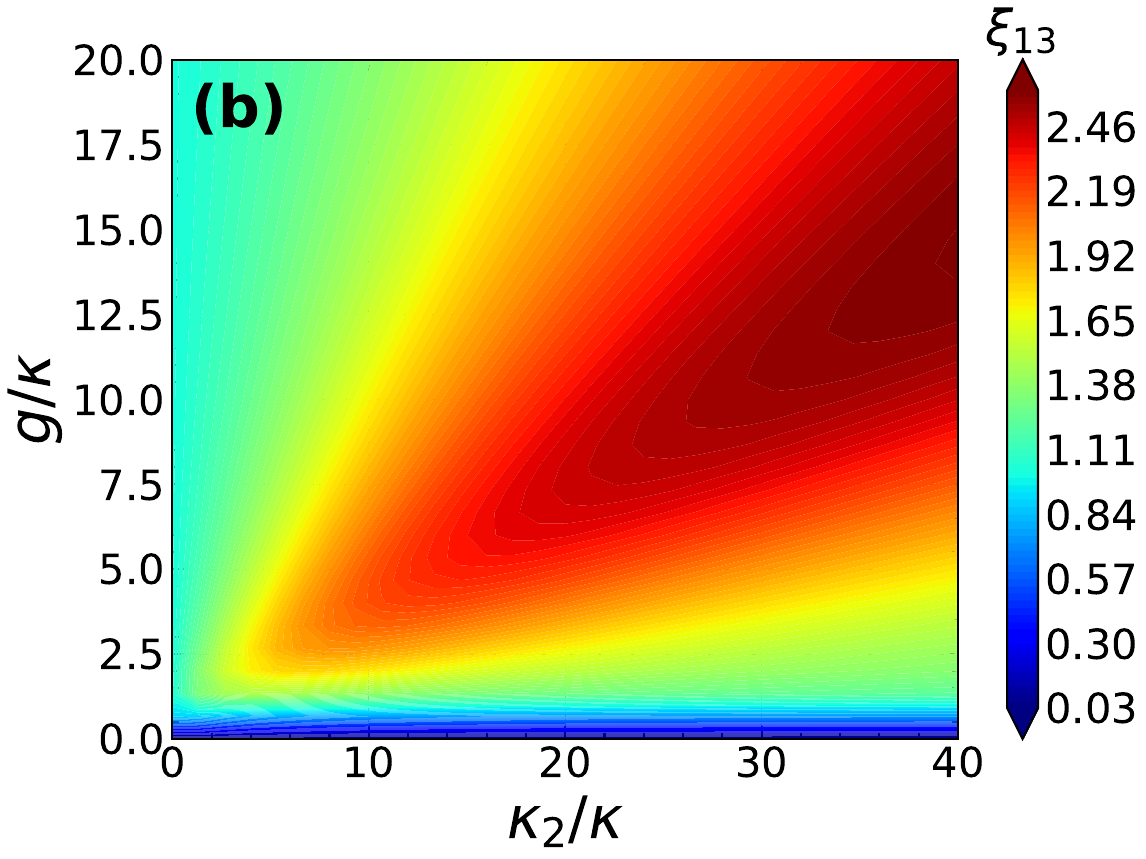}
\endminipage\hfill
\caption{(a) Stationary energy ratio $\xi_{13}$ in the $(\lambda/\kappa,\Delta/\kappa)$ plane with $\kappa_2 = 50\kappa$. (b) Stationary energy ratio $\xi_{13}$ in the $(\kappa_2/\kappa, \rm g/\kappa)$ plane. Other parameters are fixed to the reference values in Table~\ref{table1}.}
\label{fig5}
\end{figure}

We plot in Fig.~\ref{fig5}(a) the stationary energy ratio $\xi_{13}={\cal E}_{\rm B}/{\cal E}_{\rm ch}$ against the normalized OPA gain $\lambda/\kappa$ and detuning $\Delta/\kappa$, for a strongly dissipative auxiliary cavity with $\kappa_2=50\kappa$. In this case, the large auxiliary loss strongly suppresses coherent energy transfer through the intermediate cavity, thereby modifying the interference responsible for directional energy accumulation. Moreover, the contour map shows that $\xi_{13}$ is largest in the weak-parametric-drive region, particularly close to $\lambda/\kappa=0$, where values exceeding unity can be obtained for appropriate detunings. As the OPA gain increases, the stationary ratio decreases rapidly and becomes relatively insensitive to the detuning. This behavior indicates that, for very strong auxiliary dissipation, increasing the parametric gain does not necessarily improve the relative energy accumulation in the battery. However, the additional excitations generated by the OPA are increasingly accompanied by enhanced population of the charger, reducing the ratio ${\cal E}_{\rm B}/{\cal E}_{\rm ch}$. Furthermore, the weak dependence on $\Delta/\kappa$ at finite $\lambda$ further indicates that strong auxiliary damping dominates the coherent detuning-induced interference. Thus, Fig.~~\ref{fig5}(a) demonstrates that the OPA enhancement of the absolute stored energy should be distinguished from the energy ratio, particularly in the strongly dissipative regime.%

Fig.~\ref{fig5}(b) explores the steady-state ratio $\xi_{13}$ across a wide range of auxiliary damping rates $\kappa_2/\kappa$ and coupling strengths $\rm g/\kappa$. The response is strongly non-monotonic with respect to $\kappa_2$: when $\kappa_2 \to 0$, nonreciprocity is absent; as $\kappa_2$ increases into the regime $\kappa_2 \gg \{\rm g_{1},g_{2},\kappa_1, \kappa_3,|\Delta_j|\}$ ($j=1,2,3$), the auxiliary mode can be adiabatically eliminated. However, if $\kappa_2$ becomes excessively large ($\kappa_2 \gg 100\kappa$), the auxiliary cavity acts as a severe dissipation sink that absorbs photons before they can be transferred to cavity $\rm a_3$, suppressing the net energy transfer. Therefore, auxiliary dissipation acts as an essential quantum resource that must be optimized ($\kappa_2 \approx 10\kappa$) to maximize nonreciprocal energy storage.%

For $\Omega=0$, the OPA becomes the only explicit coherent energy source. For a fixed value of $\lambda$, a non-zero mode occupancy persists because the parametric amplification term continuously generates excitation pairs. This limit provides a clean way to quantify the energy supplied exclusively by the parametric amplifier.

\begin{figure}[!htb]
\minipage{0.5\textwidth}
\includegraphics[width=1\linewidth]{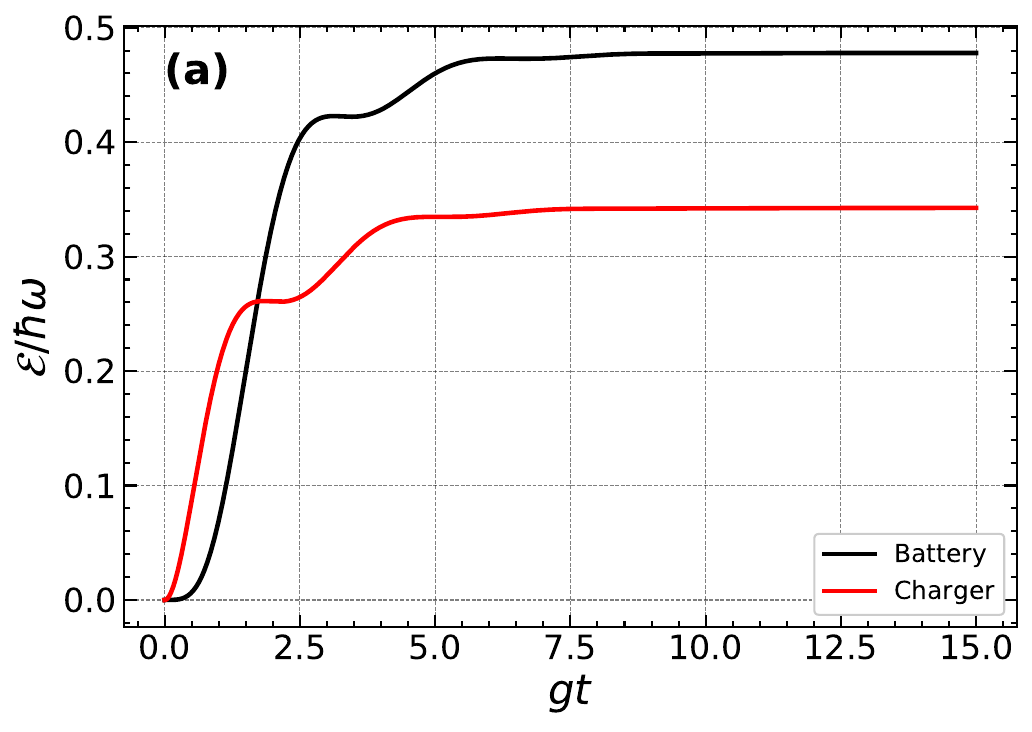}
\endminipage\hfill
\minipage{0.5\textwidth}
\includegraphics[width=1\linewidth]{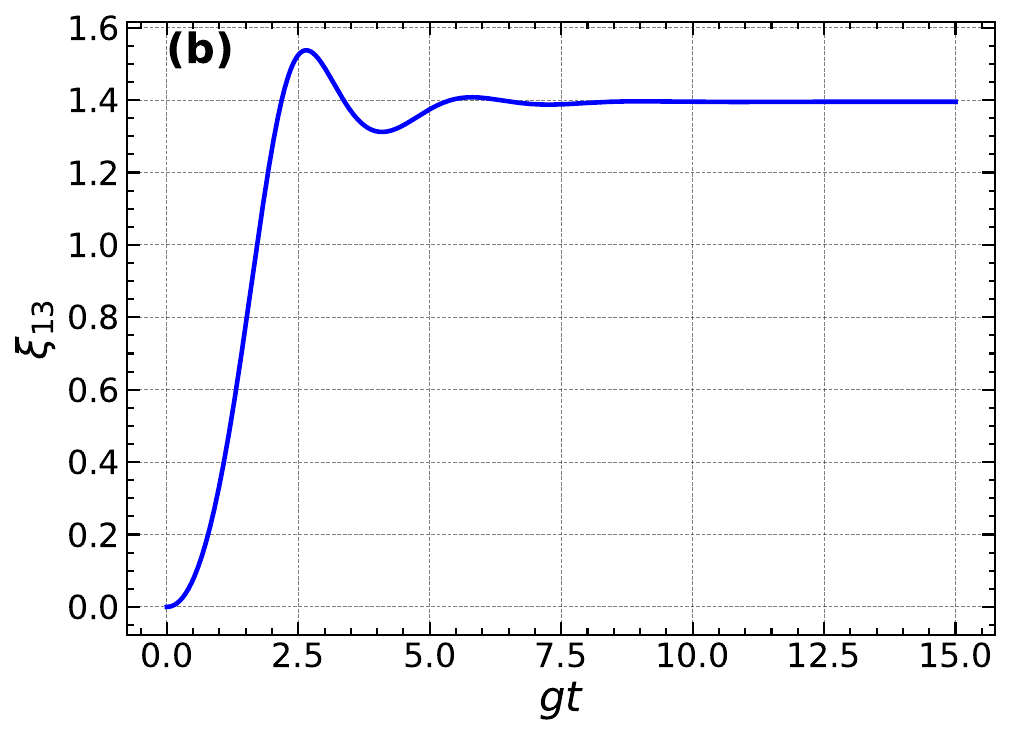}
\endminipage\hfill
\caption{Plot (a) charger and battery energies (${\cal E}_{\rm ch}(t) = \hbar\omega_1 \langle {\rm a_1}^\dagger {\rm a_1} \rangle$ and ${\cal E}_{\rm B}(t) = \hbar\omega_2 \langle {\rm a_3}^\dagger {\rm a_3} \rangle$) as functions of ${\rm g}t$ and (b) corresponding energy ratio $\xi_{13} = {\cal E}_{\rm B}/{\cal E}_{\rm ch}$, without external drive $\Omega=0$. Other parameters are fixed to the reference values in Table~\ref{table1}. }
\label{fig8}
\end{figure}

Fig.~\ref{fig8} shows the dynamical evolution of the stored energies ${\cal E}_{\rm ch}(t) = \hbar\omega_1 \langle {\rm a_1}^\dagger {\rm a_1} \rangle$ and ${\cal E}_{\rm B}(t) = \hbar\omega_2 \langle {\rm a_3}^\dagger {\rm a_3} \rangle$, together with the energy ratio $\xi_{13} = {\cal E}_{\rm B}/{\cal E}_{\rm ch}$, starting from the joint vacuum state in the absence of an external drive ($\Omega = 0$), where the parametric amplifier serves as the sole energy source ($\lambda > 0$). Moreover, under this condition, energy injection occurs purely through quantum parametric down-conversion ${\cal H}_{\text{OPA}} = \frac{i\lambda}{2}(e^{-i\phi}{\rm a_1}^{\dagger 2} - e^{i\phi}{\rm a_1}^2)$, which generates photon pairs directly from the vacuum state $|0,0,0\rangle$. We remark in panel (a) that both charger and battery modes develop finite steady--state populations, while panel (b) shows that the energy ratio $\xi_{13}$ approaches unity. This means that the excitations created in pairs in cavity ${\rm a_1}$ are successfully converted into single-mode populations and directionally transferred to cavity ${\rm a_3}$ through the loss--engineered network. This limit is particularly useful for separating energy supplied by the coherent drive from energy supplied by the parametric pump. Besides, it also emphasizes why the OPA gain should be treated as an external energetic resource rather than merely as a modification of the charger Hamiltonian. Thus, Fig.~\ref{fig8} provides a direct demonstration that the proposed charging mechanism can operate using parametric excitation alone while remaining governed by the same loss--engineered transfer network.%

\subsection{Charging power and ergotropy}  

We now examine two complementary measures of battery performance: the rate at which energy is delivered to the battery and the amount of that energy that can subsequently be extracted as useful work. The charging power and ergotropy therefore provide information that cannot be obtained from the stored energy alone.

\begin{figure}[!htb]
\minipage{0.5\textwidth}
\includegraphics[width=1\linewidth]{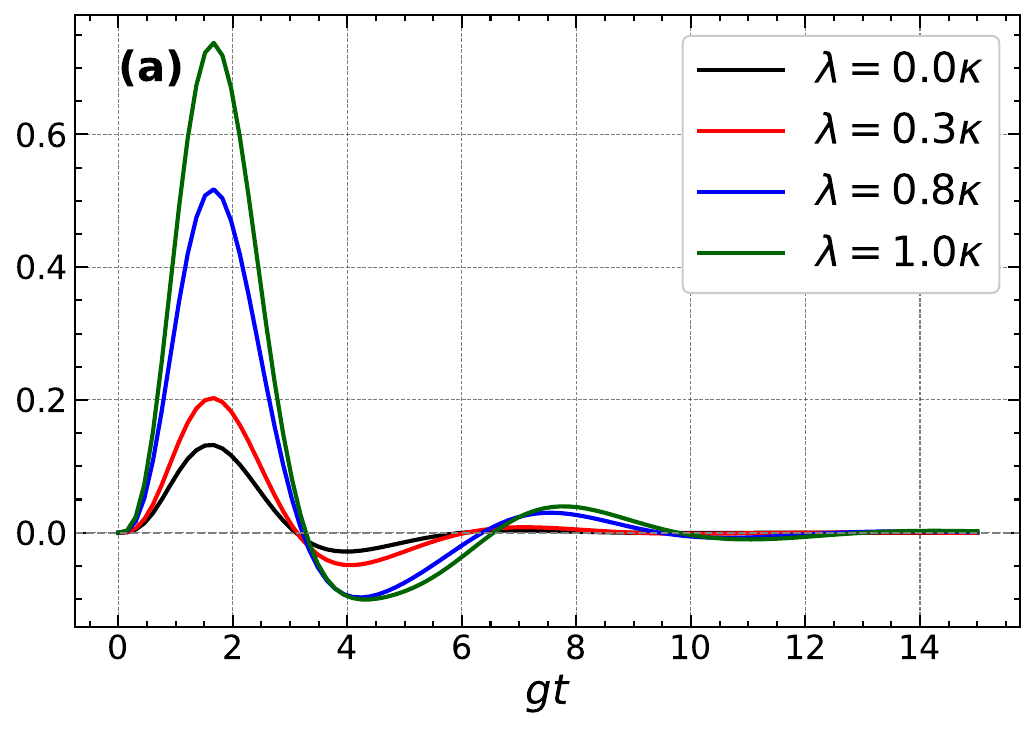}
\endminipage\hfill
\minipage{0.5\textwidth}
\includegraphics[width=1\linewidth]{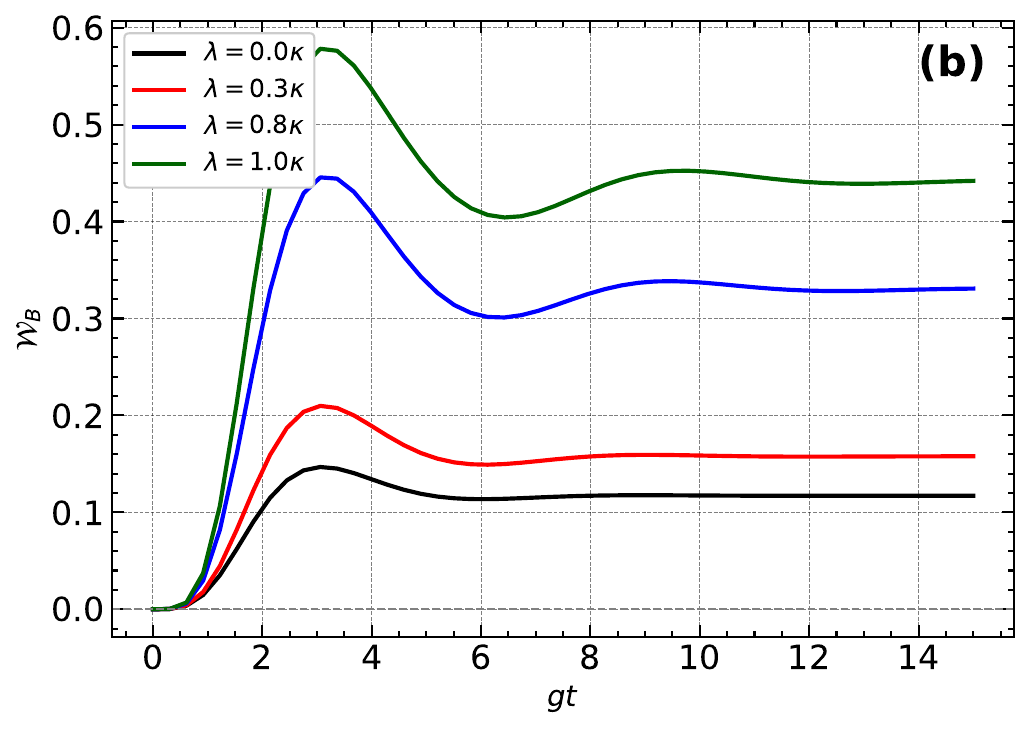}
\endminipage\hfill
\minipage{0.5\textwidth}
\includegraphics[width=1\linewidth]{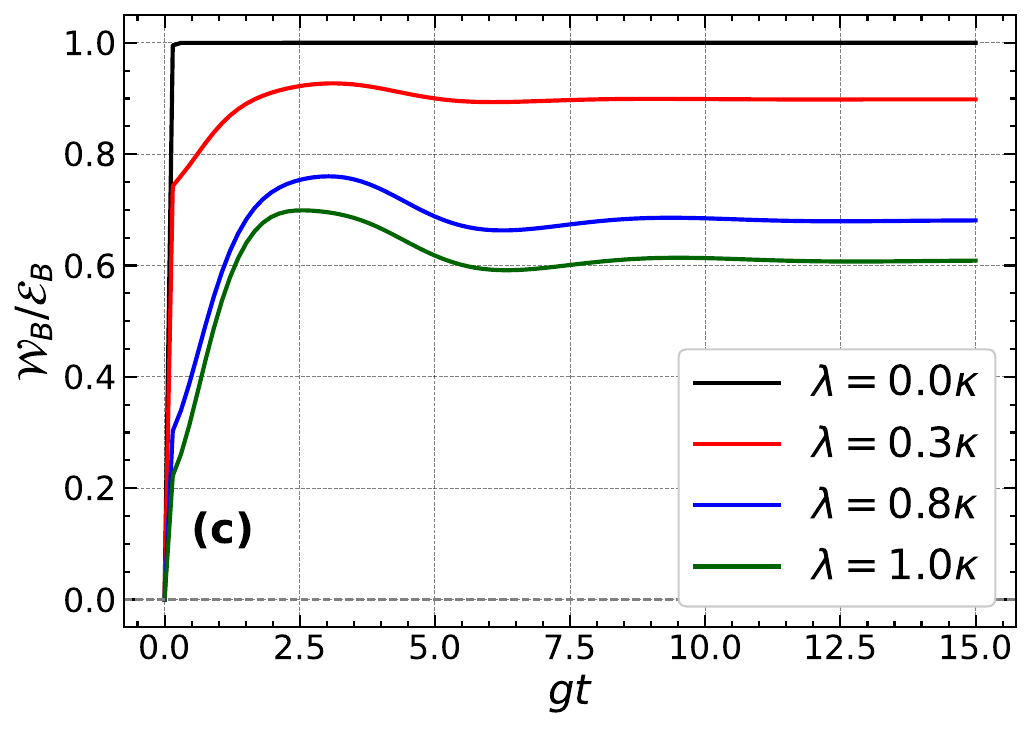}
\endminipage\hfill
\caption{Dynamical evolution of the (a) charging power ${\cal P}_{\rm B}(t)$, (b) battery ergotropy ${\cal W}_{\rm B}(t)$ and (c) ergotropy ratio ${\cal W}_{\rm B}(t)/{\cal E}_{\rm B}(t)$ for different value of the parameter OPA gain $\lambda$. Other parameters are fixed to the reference values in Table~\ref{table1}. V lam-7}
\label{fig9}
\end{figure} 

Fig.~\ref{fig9}(a) displays the dynamical evolution of the charging power ${\cal P}_{\rm B}(t)$ for parametric gain values $\lambda = 0, 0.3\kappa, 0.8\kappa, \kappa$. Starting from zero at $t=0$ for the joint vacuum state, ${\cal P}_{\rm B}(t)$ exhibits a rapid initial rise to a sharp transient peak ${\cal P}_{{\rm B},\text{max}}$ near ${\rm g}t \approx 1.8$, followed by damped oscillations toward zero as the battery approaches steady--state ($\frac{\dd {\cal E}_{\rm B}}{dt} \to 0$). Increasing the OPA gain $\lambda$ dramatically amplifies ${\cal P}_{{\rm B},\text{max}}$ (more than quadrupling its magnitude at $\lambda = \kappa$ compared to the unassisted case $\lambda = 0$). This means that the degenerate parametric amplifier acts as an active two--photon pump in the charger mode $\rm a_1$, continuously generating squeezed quadratures and correlations $\langle \rm a_1^2 \rangle$. This non--classical excitation boost creates a strong intra-cavity energy gradient that improves energy battery cavity $\rm a_3$ via both the direct path ${\rm g}_{13}e^{i\theta}$ and the indirect mediator path ${\rm g}_{12}--{\rm g}_{23}$. The subsequent negative transient dips in ${\cal P}_{\rm B}(t)$ can be explained by coherent back--flow of energy from the battery into the lossy coupled network before dynamical equilibrium is established.\\%

Fig.~\ref{fig9}(b) presents the time evolution of the battery ergotropy ${\cal W}_{\rm B}(t) = \rm{Tr}(\varrho_{\rm B} {\cal H}_{\rm B}) - \rm{Tr}(\varrho_{\rm B}^{\rm{pas}} {\cal H}_{\rm B})$, representing the maximum work extractable via arbitrary unitary transformations, where $\varrho_B^{\text{pas}}$ is the passive state. One can see that across all parametric gains $\lambda$, ${\cal W}_{\rm B}(t)$ increases monotonically during the early charging regime, reaching a maximum value around ${\rm g}t \approx 3.2$ before decreasing into a stable, non-zero stationary value of ${\cal W}_{\rm B}(t)$. Moreover, activating and increasing the OPA gain $\lambda$ significantly enhances both the peak ergotropy and the long-time steady-state ergotropy, as depicted in Fig.~\ref{fig9}(b). This mains that the photon pairs injected by the OPA retain strong non--classical coherence and quadrature squeezing that convert directly into extractable work. Furthermore, even in the presence of local cavity loss $\kappa_3$, the interference--engineered nonreciprocal coupling prevents complete energy back--flow, ensuring that the enhanced ergotropy remains securely trapped in the battery. Consequently, active parametric amplification provides a robust mechanism to scale up the absolute work storage capacity of optical cavity.\\ %

In Fig.~\ref{fig9}(c) we evaluate the ergotropy ratio ${\cal W}_{\rm B}(t)/{\cal E}_{\rm B}(t)$ over time ${\rm g}t$, quantifying the fraction of total stored battery energy that can be extracted as useful work. One can see in the absence of parametric gain ($\lambda = 0$), the battery approaches a nearly coherent displaced state, resulting in a large ergotropy-to-energy ratio. However, as the OPA gain $\lambda$ is increased, the steady--state ratio ${\cal W}_{\rm B}/{\cal E}_{\rm B}$ systematically decreases, dropping to $\approx 0.9$ for $\lambda = 0.3\kappa$, $\approx 0.7$ for $\lambda = 0.8\kappa$, and $\approx 0.6$ for $\lambda = \kappa$. One can explain this reduction arises because two--photon parametric amplification inherently introduces photon number fluctuations and mixedness alongside quadrature squeezing. Also broaden the Fock--state distribution, raising the energy baseline of the passive state $\rho_{\rm B}^{\text{pas}}$ relative to total energy ${\cal B}_{\rm B}$. From this figure the OPA gain $\lambda$ vastly increases absolute stored energy ${\cal E}_{\rm B}$ and extractable work ${\cal W}_{\rm B}$, but slightly reduces the fractional work quality ${\cal W}_{\rm B}/{\cal E}_{\rm B}$, i.e. $\rm{Tr}(\varrho_{\rm B} {\cal W}_{\rm B}) > \rm{Tr}(\varrho_{\rm B}^{\rm{pas}} {\cal W}_{\rm B})$, due to squeezing-induced quantum noise.%

\subsection{Charging power and ergotropy under pure dephasing}

We finally examine the robustness of the parametrically enhanced charging process against pure dephasing, which suppresses phase coherence without directly removing excitations from the battery. This analysis clarifies how phase noise modifies the charging power, ergotropy, and fraction of stored energy that remains extractable. The effects of pure dephasing are formally modeled by incorporating the following Lindblad term into the master equation:  $\mathcal{D}_p[\rm a^\dagger_1 a_1]\varrho = \frac{\gamma_p}{2} \left[ 2 {\rm a}_1^\dagger {\rm a}_1 \varrho {\rm a}_1^\dagger {\rm a}_1 - \{ ({\rm a}_1^\dagger {\rm a}_1)^2, \varrho \} \right]$, where $\gamma_p$ represents the pure dephasing rate associated with the cavity mode.
\begin{figure}[!htb]
\minipage{0.5\textwidth}
\includegraphics[width=1\linewidth]{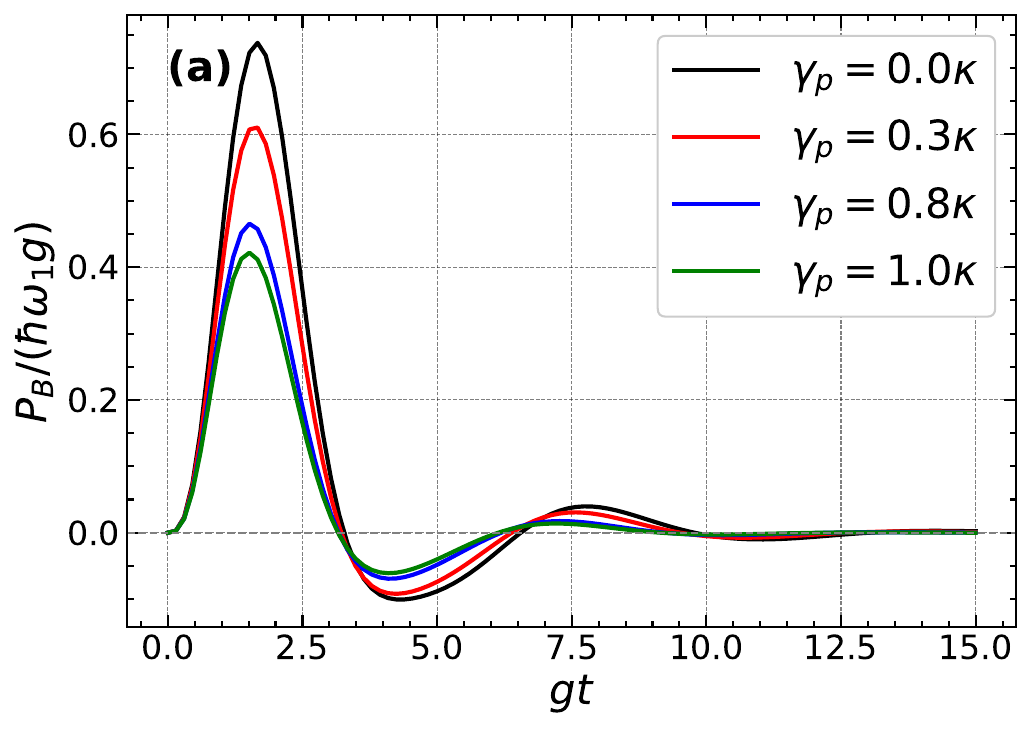}
\endminipage\hfill
\minipage{0.5\textwidth}
\includegraphics[width=1\linewidth]{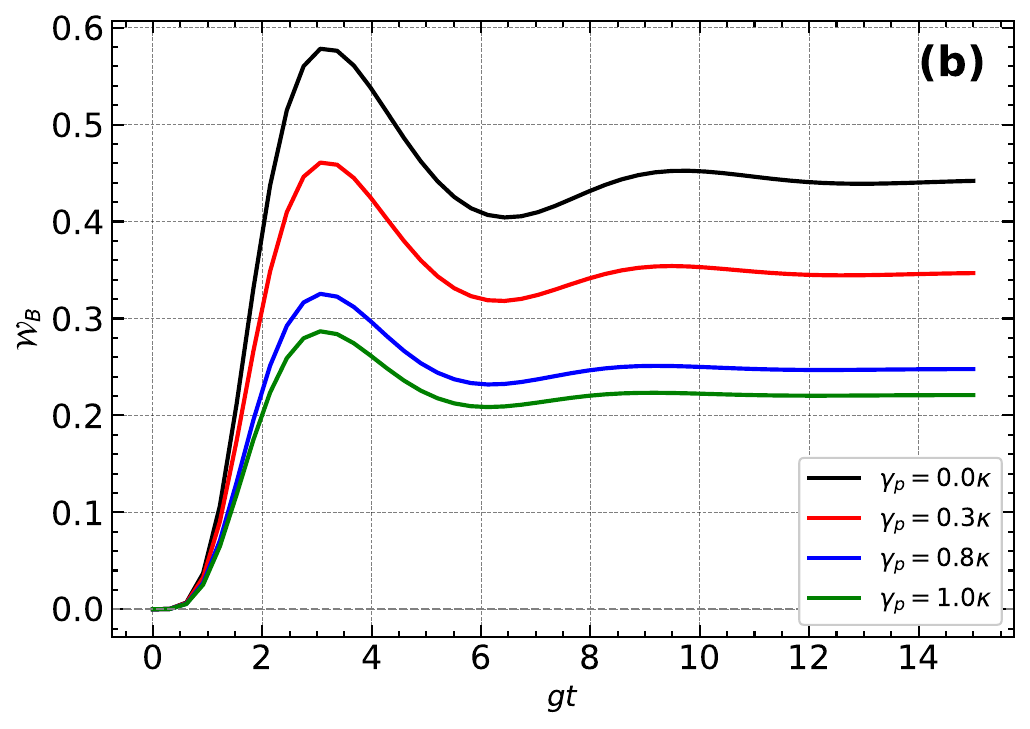}
\endminipage\hfill
\minipage{0.5\textwidth}
\includegraphics[width=1\linewidth]{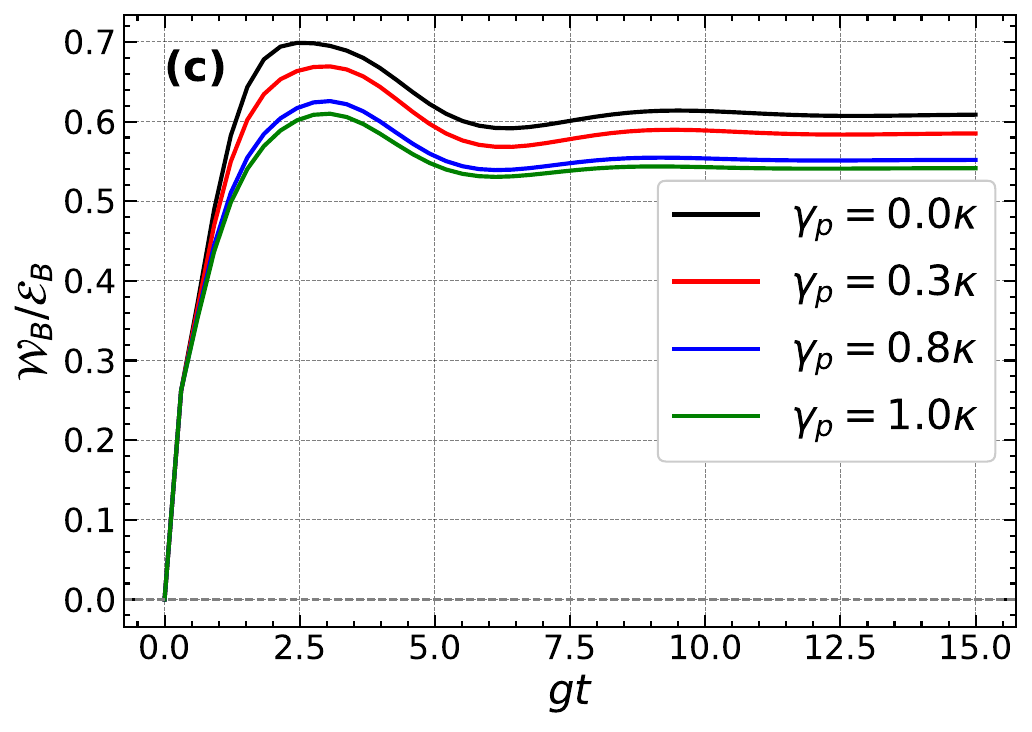}
\endminipage\hfill
\caption{Dynamical evolution of the (a) charging power ${\cal P}_{\rm B}(t)$, (b) battery ergotropy ${\cal W}_{\rm B}(t)$ and (c) ergotropy ratio ${\cal W}_{\rm B}(t)/{\cal E}_{\rm B}(t)$ for different value of the parameter of pure dephasing effect rate $\gamma_p$ with $\lambda/\kappa = 1$. Other parameters are fixed to the reference values in Table~\ref{table1}.}
\label{fig10}
\end{figure}
Fig.~\ref{fig10} evaluates the influence of pure dephasing rate $\gamma_p$ on the charging power $\mathcal{P}_{\rm B}$, ergotropy ${\cal W}_{\rm B}$, and the ergotropy ratio ${\cal W}_{\rm B}/{\cal E}_{\rm B}$). 

Fig.~\ref{fig10}(a) represents the dynamical evolution of the charging power $\mathcal{P}_{\rm B}(t)$ for various values of the pure dephasing rate $\gamma_p/\kappa$. This figure exhibits a rapid initial rise toward a sharp maximum near ${\rm g}t \approx 1.8$, followed by damped oscillations that decay toward zero in the long-time limit. For $\gamma_p = 0$, we obtain the same result as in Fig.~\ref{fig9}(a). One can see that as the pure dephasing rate $\gamma_p$ increases from $0$ to $\kappa$, the peak charging power $\mathcal{P}_{\rm B,max}$ degrade. This means that the energy transfer into the battery depends on the quantum coherence and intermode phase alignment between the charger and battery modes. Moreover, the pure dephasing causes elastic phase--randomization events without directly removing photons from the system. This loss of phase coherence degrades the constructive interference between the direct coupling ${\rm g}_{3}e^{i\theta}$ and the indirect path ${\rm g}_{1}--{\rm g}_{2}$ mediated by cavity $\rm a_2$. Consequently, the rate of energy flow into cavity $\rm a_3$ slows down, suppressing the transient power peak while preserving the underlying directional trapping.

In Fig.~\ref{fig10}(b) we present the temporal evolution of the maximum work extractable $\mathcal{W}_{\rm B}(t)$ for different pure dephasing rates $\gamma_p/\kappa$. We remark that for all value of $\gamma_p$, $\mathcal{W}_{\rm B}(t)$ increases monotonically, and reaches a maximum value at ${\rm g}t \approx 3.2$, and then settles into a stable non-zero steady--state value. Moreover, by increasing $\gamma_p$ reduces both the maximum peak ergotropy and the long--time stationary work storage capacity. Pure dephasing continuously destroys off--diagonal density matrix elements and increases decoherence, which broadens the population distribution in the Fock basis. This increased mixedness raises the energy level of the corresponding passive state $\varrho_B^{\text{pas}}$, leaving less energy available for unitary extraction. Despite this dephasing--induced degradation, the loss--engineered nonreciprocal pathway continues to prevent energy back--flow, ensuring robust, non--zero steady--state ergotropy.

In Fig.~\ref{fig10}(c) we plot the dynamical evolution of the ratio $\mathcal{W}_{\rm B}(t)/\mathcal{E}_{\rm B}(t)$, which characterizes the fraction of total stored battery energy that can be extracted as useful work. This figure shows that the ratio rapidly reaches its peak around ${\rm g}t \approx 3$ before stabilizing at steady--state. As the pure dephasing rate $\gamma_p$ increases from $0$ to $\kappa$, the steady--state ratio decreases from approximately $0.70$ down to $0.53$. This means that while the OPA injects squeezed quadratures and high excitation energy, pure dephasing acts to randomize the orientation of these squeezed states in phase space. Pure dephasing suppresses phase coherence and therfore reduces the coherent interference and anomalous correlations responsible for part of the enhanced charging performance. Because total energy $\mathcal{E}_{\rm B}$ remains relatively high while extractable work $\mathcal{E}_{\rm B}$ suffers from the increased state mixedness, the fractional work quality $\mathcal{W}_{\rm B}(t)/\mathcal{E}_{\rm B}(t)$ degrade. 

\section{Conclusion} \label{sec4}

We have theoretical enhanced ergotropy and charging power by employing degenerate OPA acting on the charger. The system consists of a driven charger cavity, a lossy auxiliary cavity, and a battery cavity. Moreover, the auxiliary loss modifies the interference between direct and indirect transfer pathways and generates directional energy flow, while the OPA supplies an additional phase--sensitive energy resource. The entire analysis was formulated directly from the Lindblad master equation. We derived the equations for first moments, populations, normal correlations, and anomalous correlations. The principal physical conclusion is that the degenerate OPA provides a controllable mechanism for increasing the energy supplied to a loss--engineered nonreciprocal battery. The OPA gain and phase provide additional control over the charging dynamics, while the auxiliary loss and coupling phase preserve the directional character of the transfer. Because the OPA itself supplies energy, however, enhanced battery energy must be distinguished from enhanced intrinsic transfer efficiency. Besides, a complete assessment should therefore report stored energy, charging power, extractable work, and the external energy supplied by the parametric pump.

\end{document}